\documentstyle[12pt,a4,psfig]{article}
\begin{document}
\def\pgf{\noindent\hangindent=0.5truein\hangafter=1}
\parindent=0mm
\parskip=3pt plus1pt minus.5pt
\sloppy
{\large \bf A PRIMORDIAL r-PROCESS?}
\vspace{0.5cm}

{\it T.\ Rauscher$^{1,2}$, 
J.H.\ Applegate$^3$, J.J.\ Cowan$^4$, F.-K.\ 
Thielemann$^{1,5}$, and M.\ Wiescher$^6$}\\
{\small $^1$ Harvard-Smithsonian Center for Astrophysics, Cambridge, MA, USA\\
$^2$ Institut f\"ur Kernchemie, Universit\"at Mainz, Mainz, 
Germany\\
$^3$ Dept. of Astronomy, Columbia University, New York, NY, USA\\
$^4$ Dept. of Physics and Astronomy, Univ. of Oklahoma, Norman, OK, 
USA\\
$^5$ Institut f\"ur theoretische Physik, Universit\"at Basel, Basel, 
Switzerland\\
$^6$ Dept. of Physics, Univ. of Notre Dame, Notre Dame, IN, USA}

%
%

\section{Introduction}

A number of possible mechanisms have been suggested
to generate density inhomogeneities in the early Universe 
which could survive until the onset of primordial 
nucleosynthesis and affect the
abundances of the isotopes produced in the big bang~\cite{mal1}. The possibility
that the changes in the abundance pattern might be drastic enough to
reconcile a closure density of baryons with primordial nucleosynthesis
constraints, or to produce a characteristic signature of a phase 
transition
in the early universe has proven exciting enough to inspire a 
considerable
amount of work on inhomogeneous nucleosynthesis. In this work
we are not concerned with how the inhomogeneities were
generated but we want to focus on the effect of such inhomogeneities on
primordial nucleosynthesis.
One of the proposed signatures of inhomogeneity, the 
synthesis of very heavy elements by neutron capture, is analyzed
for varying baryon to photon ratios $\eta$ and length scales $L$. 
A detailed discussion is published in~\cite{rau1}. Preliminary results 
can be found in~\cite{thi1}.

\section{Method}

After weak decoupling the vastly different mean free paths of protons 
and neutrons create a very proton rich environment in the initially high 
density regions, whereas the low density regions are almost entirely 
filled with diffused neutrons. Since the aim of the present 
investigation was to explore the production of heavy elements, we 
considered only the neutron rich low density zones. High density, proton 
rich, environments might produce some intermediate elements via the 
triple-alpha-reaction, but will in no case be able to produce heavy 
elements beyond iron. However, we included the effects of the (back) 
diffusion of neutrons into the proton rich zones. Using a similar 
approach as introduced in~\cite{app1,app2}, the neutron diffusive loss 
rate $\kappa$ is given by
\begin{equation}
\kappa={4.2\times 10^4 \over 
(d/a)_{cm\,MeV}}T_9^{5/4}(1+0.716T_9)^{1/2}s^{-1}
\end{equation}
in the temperature range $0.2<T_9<1$. Thus, the only open parameter in 
the neutron loss due to diffusion is the comoving length scale of 
inhomogeneities ($d/a$). Small separation lengths between high density 
zones make the neutron
leakage out of the small low density zones most effective. Large
separation lengths make the neutron leakage negligible. (For a detailed 
derivation of Eq.(1), see also~\cite{rau1}).

Our reaction network consisted of two
parts, one part for light and intermediate nuclei, the second part
being an r-process code.
For light and intermediate nuclei from neutrons and protons to krypton
($Z$=36), from stability to the neutron drip line, we made use of a
general nuclear network (of 655 nuclei) which includes neutron,
charged particle, and photon induced reaction as well as weak reactions.
(For details of the included rates see Appendix B and Tables 1, 2, A1, 
and A2 in~\cite{rau1}).

The second part was an r-process
code that determines the abundances of heavy nuclei.
This network extends up to $Z=114$ and contains all (6033) nuclei from the
so-called valley of beta-stability to the neutron-drip line (see 
also~\cite{cow1}).
The neutron capture rates were calculated with statistical model methods
and the beta decay rates were taken from~\cite{kla1},
where experimental values were not available
(see also~\cite{cow2}).
For the calculations in this paper
we also introduced (beta delayed) fission of heavy nuclei, as calculated
by Thielemann, Metzinger, and Klapdor (1983)~\cite{thi2} with the Howard 
and M\"oller
(1980)~\cite{how1} fission barriers and masses (see~\cite{cow2}).
These two networks were coupled together such that they both ran 
simultaneously
at  each time step, and  the number of neutrons produced and captured was
transmitted back and forth between them.

\section{Results and Discussion}

The choice of an initial neutron abundance of $X_n=1$ (i.e.\ only neutrons, 
which is the most favorable condition for the formation of heavy 
elements) in the low density region leads to a density ratio 
$\rho_{\mathrm{low}}/\rho_{\mathrm{b}}=1/8$~\cite{rau1}. This leaves as 
open parameters the baryon to photon ratio 
$\eta=n_b/n_{\gamma}=10^{-10}\eta_{10}$ and the comoving length scale 
($d/a$). Four sets of calculations have been performed, employing 
$\eta_{10}$ values of 416, 104, 52, and 10.4. Using the 
relation~\cite{rau1}
\begin{equation}
\Omega_b h_{50}^2=1.54 \times 10^{-2} (T_{\gamma o}/2.78K)^3
\eta_{10}\quad,
\end{equation}
with the well known present temperature of the microwave background 
$T_{\gamma o}$ and the Hubble constant $H_o=h_{50}$ $\times$ 50 km 
s$^{-1}$ Mpc$^{-1}$,
this
corresponds to possible choices of ($h_{50},\Omega_b$) being (2.5,1),
(1.3,1), (1,0.8), and (1,0.16).
The range covered in $\eta_{10}$
extends from roughly a factor of 2.2 below the lower limit to a factor 
of
13 above the upper limit for $\eta$ in the standard big bang. For each
of the $\eta$-values we considered four different cases of $d/a$:
(0) $\infty$, resulting in negligible neutron back diffusion,
(1) 10$^{7.5}$ cm MeV, (2) 10$^{6.5}$ cm MeV, and (3) 10$^{5.5}$ cm MeV.
(This corresponds to distances between nucleation sites of $\infty$, 
2700, 
270, and 27 m, respectively, at the time of the quark-hadron phase 
transition).
The resulting abundances for heavy elements are shown in Tab.\ 1.
One notices the exponential increase in r-process abundances with 
increasing $\eta$. This is due to ``fission
cycling'', whereby each of the fission fragments
can capture neutrons and finally form again heavy nuclei, which are also
prone to fission~\cite{see1}.
This is  of particular importance in environments with a long duration
of high neutron densities, and was therefore suggested as relevant to
primordial nucleosynthesis in neutron rich zones of an inhomogeneous big 
bang~\cite{app1}. In contrast to the operation of the r-process in 
explosive
stellar environments, confined to a
few seconds, this process in the neutron rich regions associated with
an inhomogeneous
big bang is only limited by the neutron half-life and can go on for an
extended period of time.
One of the remarkable features of an r-process with
fission cycling is that the production of heavy nuclei is not limited
to the r-process flow (neutron captures and beta-decays) coming from 
light
nuclei, but requires only a small amount of fissionable nuclei to be 
produced
initially. The total mass fraction of heavy nuclei is doubled with each 
fission cycle
and can thus be written as $X_r= 2^n X_{seed}$.
Here, $n$ is the number of fission cycles and $X_{seed}$ denotes the 
initial
mass fraction of heavy nuclei. The effectiveness of fission cycling can 
be seen in Fig.\ 1.
The main parameter that determines the number of fission cycles is the
rate of the r-process flux, which is a function of the location of
the r-process path with respect to the stability line, and thus dictates 
the
typical beta decay half lives. The location of the r-process path is a
function of the neutron number density $n_n$ and temperature $T$, coming 
closer to stability for decreasing
$n_n$ or increasing $T$, thus increasing cycle times and decreasing the 
number of cycles which in turn leads to smaller abundance predictions.

Since the formation of heavy elements beyond Fe and Kr is a very 
sensitive measure of $\eta$, it can be used to provide an independent 
upper limit for the product $\Omega_b H_0^2$. Fig.\ 2 shows
observational (upper) limits to the primordial creation of heavy~\cite{cow2,bee1,mat1} and 
light~\cite{mey1,kur1,rya1,dun1} element abundances to our 
results. The tightest constraints are given by the light elements 
including Li, Be, and B (see however recent doubts on the primordial
$^7$Li abundance~\cite{del1}) for which the conditions cannot differ 
much from the standard big bang.

How do changes in the reaction rates leading 
to heavy elements affect our results? Some test calculations were performed with a 
variation of the $^8$Li($\alpha$,n)$^{11}$B rate, one of the two bottle necks
toward heavy nuclei (the other one being 
$^{14}$C($\alpha$,$\gamma$)$^{18}$O). Recent experiments~\cite{mao1} 
seem to suggest that the rate used in our calculations~\cite{rau2} has to be 
increased by a factor of 3. It was found that such a change enters only 
linearly in the resulting heavy element abundances. This can be 
understood easily, as the seed production of heavy elements varies 
linearly with the Li-rate, and even for strong fission cycling this 
behavior is not changed because $X_{heavy}=X_{seed}\times 2^n$. A 
similar effect was found for the $^{18}$O(n,$\gamma$)$^{19}$O rate which 
was changed by a factor of 10 in a recent investigation~\cite{bee2}. 
Thus, the total change in heavy element abundances is a factor of 30. 
This is also shown in Fig.\ 2. (However, note that these changes are 
not included in the values given in Tab.\ 1).

Provided that density
fluctuations exist with large scale lengths in comparison to the neutron
diffusion length, the corresponding
limits for $\eta_{10}$ or $\Omega_b h_{50}^2$ change to 104 and 1.6,
respectively, at which heavy element
abundances are produced in inhomogeneous big bang models at a level
comparable to the ones seen at lowest observable metallicities.
This reduces the difference between the constraints from light and heavy
elements, although the light element constraint is still tighter.
It also underlines that not all reactions of importance are fully
explored, yet, and future changes can be expected.

{\bf Acknowledgement:} TR is supported by the Alexander 
von Humboldt foundation.

\begin{table}
\begin{center}
\begin{tabular}{rcc|c}
$\eta_{10}$&$\Omega_b h_{50}^2$&$d/a$&$>$Kr \\
\hline\hline
416&6.4&0&0.170$\times$10$^{-02}$ \\
 &&1&0.133$\times$10$^{-04}$ \\
 &&2&0.190$\times$10$^{-13}$ \\
 &&3& ---  \\
\hline
104&1.6&0&0.227$\times$10$^{-11}$ \\
 &&1&0.735$\times$10$^{-13}$ \\
 &&2& --- \\
 &&3& --- \\
\hline
52&0.8&0&0.628$\times$10$^{-15}$ \\
 &&1&0.253$\times$10$^{-16}$ \\
 &&2& --- \\
 &&3& --- \\
\hline
10.4&0.16&0& --- \\
 &&1& --- \\
 &&2& --- \\
 &&3& --- \\
\hline \hline
\end{tabular}
\end{center}
\caption{\small Mass fractions of heavy nuclei (see text).}
\end{table}
\begin{figure}
\psfig{file=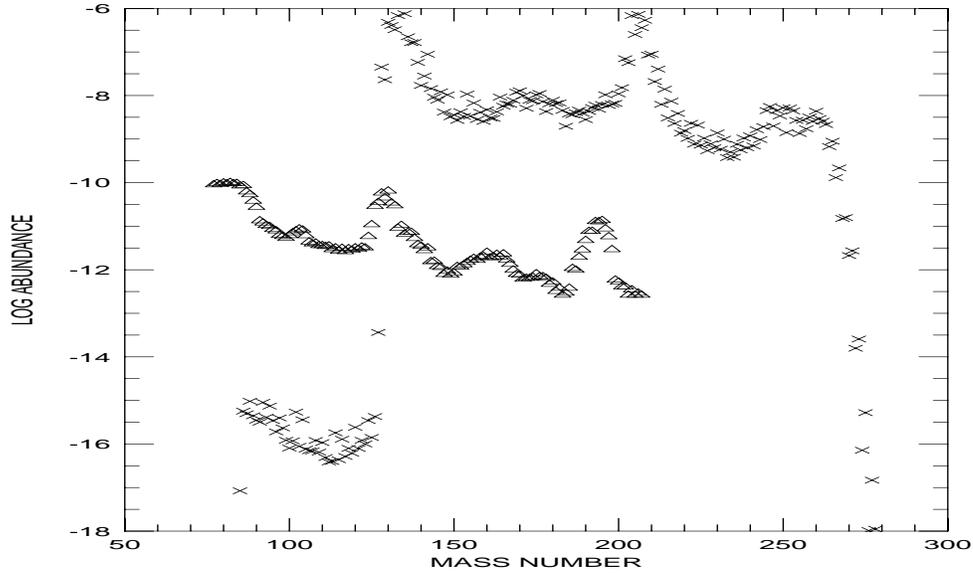,width=15.5cm,height=9cm}
\caption{\small r-process abundances for $\eta_{10}=416$ and no diffusion 
(crosses) compared to the solar abundance (triangles). Fission cycling 
has enhanced the initial 
``seed'' amount of heavy elements by 8 orders of 
magnitude. (The shift of the abundance peaks is due to the low neutron 
densities at late time [2,3]).}
\end{figure}

\begin{figure}
\psfig{file=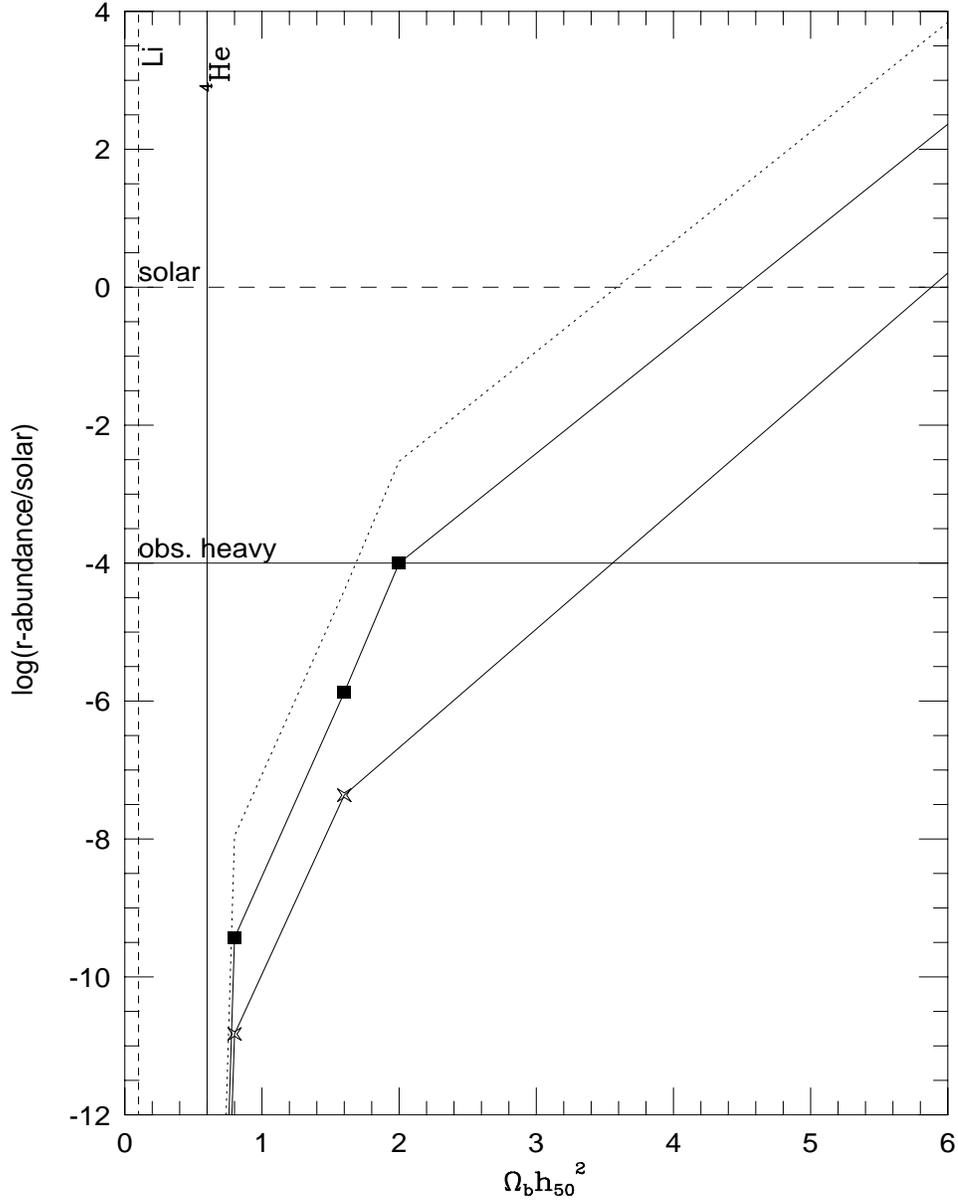,width=15.5cm,height=19cm}
\caption{\small Limits on $\Omega_b h_{50}^2$ from light and heavy element 
abundances. Abundances are normalized to solar. Shown are the results for 
different length scales of inhomogeneities, i.e.\ case 0 (full 
sq.), 1 (crosses), and 2 (open sq.). (The lines are merely drawn to guide the 
eye). Also shown is the result for the enhanced rates (dotted line). The 
horizontal solid line gives the observational upper limit for primordial 
heavy abundances. The limits on $\Omega_b h_{50}^2$ resulting from the
calculated values for the light element abundances are given by the 
vertical full and dashed lines. (See text).}
\end{figure}


\begin{thebibliography}{99}
\bibitem{mal1} R.A. Malaney and G.J. Mathews, {\it Phys.\ Rep.} {\bf 
229} (1993) 145.
\bibitem{rau1} T.\ Rauscher, J.H.\ Applegate, J.J.\ Cowan, F.-K.\ 
Thielemann, and M. Wiescher, {\it Ap.J.}, in print.
\bibitem{thi1} F.-K. Thielemann, J.H. Applegate, J.J. Cowan, and M. 
Wiescher, in {\it Nuclei in the Cosmos}, ed. H. Oberhummer (Springer 
1991), p.147.
\bibitem{app1} J.H. Applegate, {\it Phys.\ Rep.} {\bf 163} (1988) 141.
\bibitem{app2} J.H. Applegate, C.J. Hogan, and R.J. Scherrer, {\it Ap.\ 
J.} {\bf 329} (1988) 572.
\bibitem{cow1} J.J. Cowan, A.G.W. Cameron, and J.W. Truran, {\it Ap.\ 
J.} {\bf 265} (1983) 429.
\bibitem{cow2} J.J. Cowan, F.-K. Thielemann, and J.W. Truran, {\it 
Phys.\ Rep.} {\bf 208} (1991) 267.
\bibitem{kla1} H.V. Klapdor, J. Metzinger, and T. Oda, {\it At.\ Data 
Nucl.\ Data Tables} {\bf 31} (1984) 81.
\bibitem{thi2} F.-K. Thielemann, J. Metzinger, and H.V. Klapdor, {\it Z. 
Phys.\ A} {\bf 309} (1983) 301.
\bibitem{how1} W.M. Howard and P. M\"oller, {\it At.\ Data Nucl.\ Data 
Tables} {\bf 25} (1980) 219.
\bibitem{see1} P.A. Seeger, W.A. Fowler, and D.D. Clayton, {\it Ap.\ J. 
Suppl.} {\bf 97} (1965) 121.
\bibitem{bee1} T. Beers, G.W. Preston, and S.A. Shectman, {\it Astron.\ 
J.} {\bf 103} (1992) 1987.
\bibitem{mat1} G.J. Mathews, G. Bazan, and J.J. Cowan, {\it Ap.\ J.} 
{\bf 391} (1992) 719.
\bibitem{mey1} B.S. Meyer, C.R. Alcock, G.J. Mathews, and G.M. Fuller,
{\it Phys.\ Rev.\ D} {\bf 43} (1991) 1079.
\bibitem{kur1} H. Kurki-Suonio, R.A. Matzner, K.A. Olive, and D.N. 
Schramm, {\it Ap.\ J.} {\bf 353} (1990) 406.
\bibitem{rya1} S.G. Ryan, J.E. Norris, M.S. Bessel, and C.P. Deliyannis, 
{\it Ap.\ J.} {\bf 388} (1992) 184.
\bibitem{dun1} D.K. Duncan, D.L. Lambert, and D. Lemke, {\it Ap.\ J.}, 
in print.
\bibitem{del1} C.P. Deliyannis, M.H. Pinsonneault, and D.K. Duncan, {\it 
Ap.\ J.} {\bf 414} (1993) 740.
\bibitem{mao1} Z.Q. Mao, R.B. Vogelaar, A.E. Champagne, J.C. Blackmon, 
R.K. Das, K.I. Hahn, and J. Yuan, {\it Nucl.\ Phys.} {\bf A567} (1994) 
125.
\bibitem{rau2} T. Rauscher, K. Gr\"un, H. Krauss, H. Oberhummer, and E. 
Kwasniewicz, {\it Phys.\ Rev.\ C} {\bf 45} (1992) 1996.
\bibitem{bee2} H. Beer, F. K\"appeler, and M. Wiescher, in {\it Capture 
Gamma-Ray Spectroscopy}, ed.\ J. Kern (Bristol: IOP), in print.
\end{thebibliography}
\end{document}